\documentclass[aps,pra,reprint,amsmath,amssymb,superscriptaddress,nofootinbib]{revtex4-2}

\usepackage{bm}
\usepackage{mathtools}
\usepackage{amsthm}
\usepackage{microtype}
\usepackage{hyperref}
\usepackage{braket}

\hypersetup{colorlinks=true,citecolor=blue,linkcolor=blue,urlcolor=blue}

\newcommand{\Hil}{\mathcal{H}}
\newcommand{\Lin}{\mathcal{L}}
\newcommand{\States}{\mathcal{S}}
\newcommand{\Effects}{\mathcal{E}}

\newcommand{\id}{\mathcal{I}}
\newcommand{\Tr}{\operatorname{tr}}

\newcommand{\EP}{\mathrm{EP}}

\newcommand{\obsA}{\mathbf{A}}

\begin{document}

\title{Operational signatures of exceptional quantum channels}

\author{P. A. Brand\~ao}
\affiliation{Instituto de F\'isica, Universidade Federal de Alagoas, 57072-970 Macei\'o, Alagoas, Brazil}

\date{\today}

\begin{abstract}
Exceptional points (EPs) are usually introduced as defective degeneracies of non-Hermitian Hamiltonians and, in fully open quantum dynamics, of Liouvillian generators. Here we formulate an operational theory of exceptional points that does not invoke a Hamiltonian or a master equation. The primitive dynamical object is a completely positive trace-preserving (CPTP) quantum channel $\Phi$, applied an integer number $n$ of times, while physical data are probabilities $p_n^{\obsA}(x|\rho)$ associated with states $\rho$ and observables $\obsA$ having measurements results $x$. We define exceptional channels through nontrivial Jordan chains of $\Phi$ and show that a Jordan block of order $m$ generates discrete polynomial-exponential terms with maximal degree $m-1$. This places the physics of exceptional points in contact with what can be directly measured in experiments.
\end{abstract}

\maketitle

\section{Introduction}

The modern study of non-Hermitian physics was strongly influenced by the observation of Bender and Boettcher that non-Hermitian Hamiltonians with parity-time ($\mathcal{PT}$) symmetry can possess entirely real spectra in appropriate parameter regimes \cite{BenderBoettcher1998}. Exceptional points (EPs), at which eigenvalues and eigenvectors coalesce and the relevant operator becomes defective, subsequently became central objects in non-Hermitian physics \cite{Heiss2012,ElGanainy2018,Ashida2020,MiriAlu2019,Ozdemir2019}. Their consequences have been explored extensively in photonics and in non-Hermitian topology, with recent reviews emphasizing higher-order EPs, topological phenomena, sensing, and nanoscale implementations \cite{Bergholtz2021,LiEtAl2023}. In quantum systems the interpretation of such singularities requires care, because an effective non-Hermitian Hamiltonian generally describes conditional or reduced dynamics rather than the most general physical transformation of a density operator.

A fully open Markovian quantum system is more naturally described by a Liouvillian generator. This motivated the theory of Liouvillian exceptional points, in which defective degeneracies occur in the spectrum of the Liouvillian superoperator rather than in that of an effective Hamiltonian \cite{Minganti2019,Minganti2020,Khandelwal2021,Chen2022,SunYi2024}. Exceptional quantum dynamics has been investigated experimentally in dissipative superconducting qubits through state tomography and nonequilibrium observables \cite{Naghiloo2019,Erdamar2024}, while Liouvillian EPs have been reconstructed in quantum processors \cite{Abo2024}. More recently, non-Markovian quantum EPs have been observed in circuit-QED settings \cite{Zhang2025}, and theoretical studies have emphasized that quantum jumps can change the location, existence, and even the order of spectral degeneracies relative to effective non-Hermitian Hamiltonian descriptions \cite{Kopciuch2025}. These developments reinforce the importance of formulating exceptional structures at the level of the full quantum evolution.

There is, however, no fundamental reason to restrict an open quantum evolution to a continuous-time semigroup or to assume the existence of a time-local generator. The most general deterministic transformation of a finite-dimensional quantum system is a completely positive trace-preserving (CPTP) map, i.e., a quantum channel \cite{Choi1975,Kraus1983,NielsenChuang2010,HeinosaariZiman2012}. Recent work has demonstrated exceptional points directly in quantum channels, including interpolated single-qubit channels with experimentally observed second-order EPs and theoretically demonstrated third-order EPs \cite{WongZengLi2026}, and discrete CPTP brickwork circuits exhibiting exceptional spectral structure \cite{PopkovSalerno2026}. These developments suggest that exceptional-point physics can be formulated without taking a continuous-time limit.

The purpose of the present work is to push this viewpoint further and formulate exceptional points in explicitly operational terms. We regard preparations, transformations, and measurements as the primitive ingredients of an experiment. A state $\rho$ is prepared, the same quantum channel $\Phi$ is applied $n$ times, and an observable $\obsA$ having some outome $x$ is measured. The directly accessible object is therefore not an eigenvalue of a Hamiltonian or a generator, but the probability given by the Born rule:
\begin{equation}
	p_n^{\obsA}(x|\rho,\Phi)
	=\Tr\!\left[\obsA(x)\Phi^n(\rho)\right].
	\label{eq:primitive_probability}
\end{equation}
Our central question is: \emph{which features of an exceptional point are visible at the level of such probabilities, and which states and observables are required to reveal them?} The framework is deliberately independent of $\mathcal{PT}$ symmetry. It is also independent of a Lindblad equation, a Liouvillian generator, and a continuous time parameter. A control parameter may be introduced to move through a family of channels and locate a defective point, but the dynamics itself consists only of the discrete sequence
\begin{equation}
	\rho_n=\Phi^n(\rho_0),\qquad n=0,1,2,\ldots.
\end{equation}
This makes the formulation compatible with collision models and repeated-interaction schemes \cite{AttalPautrat2006,Ciccarello2022}, as well as repeated quantum operations, noisy gate sequences, stroboscopic processes, and general quantum circuits.

Section \ref{sec:operational} briefly reviews the operational approach to quantum theory and fix notation. Section \ref{sec:exceptional_channels} defines the notion of exceptional channels and discusses the Jordan chain for superoperators. In section \ref{sec:observables}, we connect the exceptional points with observables and preparation procedures. Section \ref{sec:qubit} is devoted to a complete discussion of an analytical model where all the previous discussion is applied and in section \ref{sec:conclusions} we provide our conclusions.

\section{Operational framework}
\label{sec:operational}

We first recall the basic definitions of the operational approach to quantum theory and fix notation.  We use the terminology of Ref. \cite{HeinosaariZiman2012}. Let $\Hil_d$ be a $d$-dimensional complex Hilbert space. The set of quantum states is
\begin{equation}
    \States(\Hil_d)=\{\rho\in\Lin(\Hil_d):\rho\ge 0,\;\Tr\rho=1\},
\end{equation}
where $\Lin(\Hil_d)$ is the set of linear operators acting on $\mathcal{H}_d$. The set of effects is defined by 
\begin{equation}
	\Effects(\Hil_d) = \{ E \in \Lin(\Hil_d) : O \leq E \leq I \},
\end{equation}
where $O$ is the null operator and $I$ the identity. Recall that $E\geq O$ means that $\langle \psi | E\psi\rangle \geq 0$ for every $\psi \in \Hil_d$. If this condition is satisfied, the operator $E$ is called positive. The notation $B\geq A$ means that $B-A$ is a positive operator.

We consider only experiments having a finite number of outcomes. Let $\Omega_A = \{ \omega_1,\cdots,\omega_n\}$ be a finite outcome set and $\mathcal{F}_A = 2^{\Omega_A}$ the power set representing its $\sigma$-algebra. An observable $\mathbf{A}$ is a map
\begin{equation}
    \obsA:\mathcal{F}_A\to\Effects(\Hil_d),
\end{equation}
satisfying $\obsA(\Omega_A) = I$, $\obsA(\varnothing) = O$ and
\begin{equation}\label{unionOBS}
    \obsA\left( \bigcup_{n=1}^{\infty}X_n\right) = \sum_{n=1}^{\infty}\obsA(X_n), 
\end{equation}
where $X_n\cap X_m = \varnothing$ ($n\neq m$) for all $X_n,X_m\in\mathcal{F}_A$. In this case, it is sufficient to consider the effects $E_j$ that are mapped by the observable on the elementary events $\{\omega_j\}$, that is, $E_j = \obsA(\{\omega_j\})$. The condition $\obsA(\Omega_A) = I$ translates to $$\sum_{j=1}^n E_j = I$$ which is the usual formulation of a POVM (positive operator-valued measure) found in the quantum information literature. An observable is thus a POVM. The probability of outcome $X\in\mathcal{F}_A$ when the system is prepared in state $\rho\in\mathcal{S}(\mathcal{H}_d)$ is given by the Born rule:
\begin{equation}\label{bornrule}
	p(X)  = \Tr\left[\obsA(X)\rho\right].
\end{equation}

Before some measurement is completed, one may be interested in manipulating the quantum system. A quantum channel $\Phi$ is a linear map
\begin{equation}
    \Phi:\Lin(\Hil_d)\to\Lin(\Hil_d)
\end{equation}
that is completely positive and trace preserving (CPTP). A channel admits a Kraus representation
\begin{equation}
    \Phi(\rho)=\sum_\mu K_\mu \rho K_\mu^\dagger,
    \qquad
    \sum_\mu K_\mu^\dagger K_\mu=I,
    \label{eq:kraus}
\end{equation}
where $K_{\mu}\in\Lin(\Hil_d)$ are Kraus operators. The theoretical protocol considered throughout this work has three stages:
\begin{equation}
\rho
\quad\longrightarrow\quad
\underbrace{\Phi\circ\cdots\circ\Phi}_{n\ \mathrm{applications}}
\quad\longrightarrow\quad
\text{observable }\obsA .
\end{equation}
A system is initially prepared in state $\rho$ and is submitted to a series of $n$ composite transformations by the same quantum channel $\Phi$. At the end of the process, a measurement is performed by the observable $\obsA$ and one obtains a probability distribution whose statistics are given by Eq.~\eqref{bornrule}.

\section{Exceptional quantum channels}
\label{sec:exceptional_channels}

We define an exceptional quantum channel $\Phi$ as a superoperator $\Phi : \mathcal{L}(\mathcal{H}_d)\rightarrow\mathcal{L}(\mathcal{H}_d)$ possessing at least one nontrivial Jordan block. A Jordan chain of length $m$ associated with a fixed eigenvalue $\lambda$ of a quantum channel $\Phi$ is a sequence of nonzero operators
\begin{equation}
    X_0,X_1,\ldots,X_{m-1} \in \Lin(\Hil_d)
\end{equation}
satisfying
\begin{subequations}
\label{eq:jordan_chain}
\begin{align}
    (\Phi-\lambda\id)(X_0)&=0,\\
    (\Phi-\lambda\id)(X_j)&=X_{j-1},\qquad j=1,\ldots,m-1,
\end{align}
\end{subequations}
where $\id$ is the (super)identity map acting on elements in $\Lin(\Hil_d)$. The eigenoperator $X_0$ and the generalized eigenoperators $X_j$ ($j\geq1$), are not required to be states or effects. The number $m$ is called the order of the EP $\lambda$.

Let $X_0,\ldots,X_{m-1} \in\Lin(\Hil_d)$ be a Jordan chain of $\Phi$ with eigenvalue $\lambda$. Then, for every $j=0,\ldots,m-1$ and integer $n\ge0$, we have
\begin{equation}
    \Phi^n(X_j)
    =\sum_{k=0}^{\min(j,n)}
    \binom{n}{k}\lambda^{n-k}X_{j-k},
    \label{eq:jordan_iteration}
\end{equation}
with the usual convention for the nilpotent case $\lambda=0$. To see this, consider the invariant subspace generated by the chain and write $\Phi=\lambda\id+N$, $NX_j=X_{j-1}$ and $N^m=0$. Since $\lambda\id$ and $N$ commute,
\begin{equation}
    \Phi^n=(\lambda\id+N)^n
    =\sum_{k=0}^{n}\binom{n}{k}\lambda^{n-k}N^k.
\end{equation}
Acting on $X_j$ truncates the sum at $k=j$ and yields Eq.~\eqref{eq:jordan_iteration}. For a second-order ($m=2$) EP,
\begin{equation}
    \Phi^n(X_1)=\lambda^n X_1+n\lambda^{n-1}X_0.
    \label{eq:ep2_signature}
\end{equation}
For a third-order ($m=3$) EP,
\begin{equation}
\begin{split}
    \Phi^n(X_2)=&\lambda^nX_2
    +n\lambda^{n-1}X_1\\
    &+\frac{n(n-1)}{2}\lambda^{n-2}X_0.
\end{split}
\end{equation}
Thus a block of order $m$ produces a polynomial prefactor of maximal degree $m-1$ multiplying the exponential-in-$n$ mode $\lambda^n$.

It is well known that every eigenvalue of a finite-dimensional quantum channel lies in the closed unit disk. Moreover, the algebraic and geometric multiplicity of the eigenvalues satisfying $|\lambda|=1$ are equal  \cite{EvansHoeghKrohn1978,HansonRouzeFranca2020}. Therefore, every Jordan block associated with $|\lambda|=1$ has size one. Consequently, every nontrivial exceptional eigenvalue of a quantum channel satisfies $|\lambda_{\EP}|<1$. This has a direct physical meaning. Exceptional structure in a deterministic CPTP channel cannot live in a nondecaying stationary or asymptotically periodic Jordan sector. It is necessarily encoded in transient modes that decay under repeated channel application. 

\section{Visible and blind Observables}
\label{sec:observables}


A Jordan block is not directly measured. What is measured is the classical information encoded in the probabilities given by the Born rule \eqref{bornrule}. Suppose, for simplicity, that the quantum state $\rho\in\mathcal{S}(\mathcal{H}_d)$ under investigation, relevant to a specific experiment, contains a component $X_j$ along a generalized eigenoperator of the quantum channel $\Phi$. Then, the probability of observing outcome $x\in2^{\Omega_A}$ when measuring the observable $\mathbf{A}$ would have a term of the of form
\begin{equation}
\begin{split}
    \Tr[\mathbf{A}(x)\Phi^n(X_j)]
    =\sum_{k=0}^{j}
    \binom{n}{k}\lambda^{n-k}
    \Tr[\mathbf{A}(x)X_{j-k}].
    \label{eq:probability_jordan}
\end{split}
\end{equation}
The polynomial terms therefore survive in observed probabilities only if the measured effect $\mathbf{A}(x)$ has nonzero overlap with the corresponding Jordan directions. For example, if the eigenvalue $\lambda$ is of second-order nature,
\begin{align*}
    \Tr[\mathbf{A}(x)\Phi^n(X_1)] =\lambda^n&\Tr[\mathbf{A}(x)X_1] \\
    &+n\lambda^{n-1}\Tr[\mathbf{A}(x)X_0].
    \label{eq:ep2_probability}
\end{align*}
Hence the experimental signature of the exceptional term disappears for outcome $x$ whenever $\Tr[\mathbf{A}(x)X_0]=0.$ This motivates an explicitly operational distinction between the algebraic existence of a Jordan block and its visibility to a given measurement.

Let $\Phi$ have a Jordan block associated with $\lambda$. The block is \emph{visible} to a preparation $\rho$ and an observable $\mathbf{A}$ if at least one outcome probability $p_n^{\mathbf{A}}(x|\rho,\Phi)$ contains a nonzero polynomial factor in $n$ generated by that block. If all such factors vanish for every $x\in2^{\Omega_A}$, the observable is blind to that exceptional component for the chosen preparation.

\section{Example: qubit model}
\label{sec:qubit}

As a concrete example of the previous discussion, consider a two-dimensional Hilbert space $\mathcal{H}_2$ having orthonormal basis $\{ \ket{0}, \ket{1}\}$.  We denote the Pauli operators by $\sigma_j$ ($j=x,y,z$) and the set $\{ I, \sigma_x, \sigma_y, \sigma_z\}$ form a basis in $\Lin(\Hil_2)$. Consider $0 < \lambda < 1$ and define the Kraus operators
\begin{equation}
	K_0 = \sqrt{\frac{1+\lambda^2}{2}}U_{\theta}, \quad K_1 = \sqrt{\frac{1-\lambda^2}{2}}U_{\theta}\sigma_z,
\end{equation}
where $U_{\theta} = e^{-i\theta \sigma_y /2}$ and $\theta$ an arbitrary real parameter. It is easy to check that $K_0^{\dag}K_0 + K_1^{\dag}K_1 = I$. Therefore, the superoperator $$\Phi_{\theta,\lambda} (\rho) = K_0\rho K_{0}^{\dag} + K_1\rho K_{1}^{\dag}$$ is CPTP and represents a quantum channel. In the  Pauli basis, $\Phi_{\theta,\lambda}$ is represented by the $4\times4$ matrix
\begin{equation}
	[\Phi] = \begin{pmatrix}
		1 & 0 & 0 & 0\\
		0 & \lambda^2\cos\theta & 0 & \sin\theta \\
		0 & 0 & \lambda^2 & 0 \\
		0 & -\lambda^2\sin\theta & 0 & \cos\theta
	\end{pmatrix}.
\end{equation}
It is evident that the identity $I$ and $\sigma_y$ are eigenoperators of the channel with eigenvalues $1$ and $\lambda^2$.

In the subspace $\text{span}\{ \sigma_x, \sigma_z \}$, the channel is represented by the $2\times2$ matrix
\begin{equation}\label{submatrix}
	M = \begin{pmatrix}
		\lambda^2\cos\theta & \sin\theta \\
		-\lambda^2\sin\theta & \cos\theta
	\end{pmatrix}
\end{equation}
whose eigenvalues are given by
\begin{equation}
	\mu_{\pm} = \frac{(1+\lambda^2)\cos\theta \pm \sqrt{(1 + \lambda^2)^2\cos^2\theta - 4\lambda^2}}{2}.
\end{equation}
Both eigenvalues coalesce when $(1 + \lambda^2)^2\cos^2\theta - 4\lambda^2 = 0$. By choosing the positive branch, we can write
\begin{align}
	\cos\theta_{EP} &= \frac{2\lambda}{1+\lambda^2}, \\
	\sin\theta_{EP} &= \frac{1-\lambda^2}{1+\lambda^2}.
\end{align}
In this case, $\mu_+ = \mu_- = \lambda$ and the eigenvalues of $\Phi_{EP}$ are $\{1,\lambda^2,\lambda,\lambda\}$. To demonstrate that the eigenvalue $\lambda$ is defective, that its, that $\Phi_{EP}$ is not diagonalizable, write the submatrix \eqref{submatrix} at the exceptional point, $M_{EP}$, to obtain
\begin{equation}
	M_{EP} - \lambda I = \frac{1-\lambda^2}{1+\lambda^2}\begin{pmatrix}
		-\lambda & 1\\
		-\lambda^2 & \lambda
	\end{pmatrix}.
\end{equation}
Then, it is easy to check that $(M_{EP} - \lambda I)^2 = 0$. Therefore, $M_{EP} - \lambda I$ is second-order nilpotent and the exceptional channel is of second-order nature.

The Jordan chain is generated by defining $X_0 = \sigma_x + \lambda \sigma_z$ and noting that (at the exceptional point) $$(\Phi_{EP} - \lambda \mathcal{I})X_0 = 0.$$ Therefore, $X_0$ is an eigenoperator having eigenvalue $\lambda$. Now choose $$X_1 = \frac{1+\lambda^2}{1-\lambda^2}\sigma_z$$ and notice that $$(\Phi_{EP} - \lambda \mathcal{I})X_1 = X_0.$$ In the basis $\{I,\sigma_y,X_0,X_1\}$, the channel is represented by the matrix
\begin{equation}
	[\Phi_{EP}] = \begin{pmatrix}
		1 & 0 & 0 & 0 \\
		0 & \lambda^2 & 0 & 0 \\
		0 & 0 & \lambda & 1 \\
		0 & 0 & 0 & \lambda
	\end{pmatrix}
\end{equation}
and the Jordan block appears explicitly. If the channel $\Phi_{EP}$ is applied $n$ times on the operator $X_1$, it displays the polynomial signature
\begin{equation}
	\Phi_{EP}^n (X_1) = \lambda^n X_1 + n\lambda^{n-1}X_0.
\end{equation}

Consider now a qubit prepared in the state $$\rho_0 = \frac{1}{2}(I + x_0 \sigma_x + y_0\sigma_y + z_0\sigma_z),$$ where $x_0^2 + y_0^2 + z_0^2 \leq 1$ and $(x_0,y_0,z_0)$ is the Bloch vector. After $n$ interactions with the channel $\Phi_{EP}$ we obtain the ouput state
\begin{equation}
	\rho_n = \Phi_{EP}^n(\rho_0) = \frac{1}{2}(I + x_n\sigma_x + y_n\sigma_y + z_n\sigma_z),
\end{equation}
where $(x_n,y_n,z_n)$ is the output Bloch vector. It is straightforward to demonstrate that
\begin{align}
	x_n &= \lambda^n x_0 + n\lambda^{n-1}\frac{1-\lambda^2}{1+\lambda^2}(z_0  - \lambda x_0), \\
	y_n &= \lambda^{2n}y_0, \\
	z_n &=  \lambda^n z_0 + n\lambda^{n}\frac{1-\lambda^2}{1+\lambda^2}(z_0 - \lambda x_0).
\end{align}

Suppose that a measurement is now performed on the transformed state $\rho_n$ by the sharp observable $\mathbf{A}:2^{\Omega_A} \rightarrow \mathcal{E}(\mathcal{H}_2)$ having outcome space $\Omega_A = \{+,-\}$. The effects for the elementary events are given by
\begin{align*}
	\mathbf{A}(\{+\}) &= \frac{1}{2}(I + \vec{m}\cdot\vec{\sigma}),\\
	\mathbf{A}(\{-\}) &= \frac{1}{2}(I - \vec{m}\cdot\vec{\sigma}),
\end{align*}
where $\vec{m} = (m_x,m_y,m_z)$ is a unit vector ($m_x^2 + m_y^2 + m_z^2 = 1$) with real components and $\vec \sigma = (\sigma_x,\sigma_y,\sigma_z)$. By sharp we mean that the elementary effects $\mathbf{A}(\{\pm\})$ are orthogonal rank-one projectors ($\mathbf{A}$ is a PVM).

After $n$ applications of the channel $\Phi$, and several experimental runs, we may ask for the probabilities $p_+$ and $p_-$ of observing the events $\{+\}$ and $\{ -\}$, respectively. They are given by
\begin{align}
	p_+(n)& = \text{tr}\left[ \rho_n \mathbf{A}(\{+\}) \right], \nonumber\\
		  &= \frac{1}{2}\left[ 1 + \lambda^n(m_xx_0 + m_zz_0) + m_y\lambda^{2n}y_0 \right. \nonumber\\
		 &\left. + \frac{n\lambda^{n-1}(1-\lambda^2)(z_0 - \lambda x_0)(m_x+\lambda m_z)}{1+\lambda^2}  \right] \label{pplus}
\end{align}
and $p_-(n) = 1 - p_+(n)$. This formidable equation has all the experimentally accessible signatures of the exceptional channel. It is the last term in Eq. \eqref{pplus} that contains the Jordan block signature and therefore let us assume that $z_0 - \lambda x_0 \neq 0$ and consider the initial state preparation $$\rho_0 = \frac{1}{2}\left(I + \frac{\sigma_y + \sigma_z}{\sqrt{2}}\right)$$ having Bloch vector $(0,1,1)/\sqrt{2}$. The other relevant term which must not vanish if one is to observe exceptional signatures is $m_x + \lambda m_z$. Let us consider three measurement directions given by 
\begin{align*}
	\vec{m}_a &= (1,0,0), \\
	\vec{m}_b &= (0,1,0), \\
	\vec m_c &= \frac{(-\lambda, 0, 1)}{\sqrt{1 + \lambda^2}}.
\end{align*} 
We expect that a measurement in the direction $\vec m_a$ to reveal the exceptional order structure of the channel but not for $\vec m_b$ or $\vec m_c$. This last case is interesting because it shows that the absence of exceptional signatures do not occur simply because we are measuring an uncoupled sector, as in the case of $\sigma_y$. The observable $\mathbf{A}$ constructed to measure the direction $\vec m_c$ lives in the same plane of the Jordan block.

\begin{figure}[hbt!]
	\centering
	\includegraphics[width=0.45\textwidth]{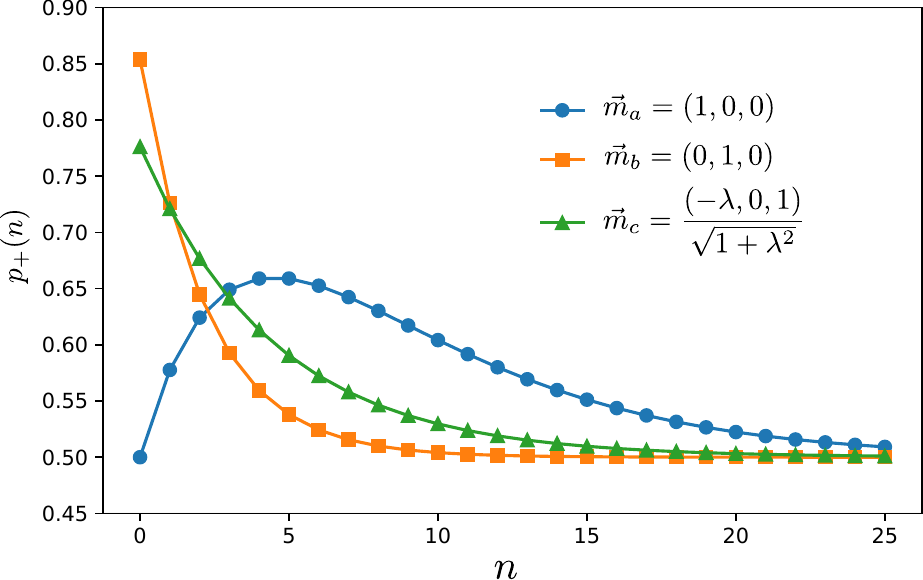}
	\caption{Probability $p_{+}(n)$ of obtaining the outcome $+$ after $n$ successive applications of the exceptional channel $\Phi_{\mathrm{EP}}$, for $\lambda=0.8$ and the initial state $\rho_0=\frac{1}{2}\left(I+\frac{\sigma_y+\sigma_z}{\sqrt{2}}\right)$. Three sharp binary observables are considered, with measurement directions	$\vec{m}_a=(1,0,0)$, $\vec{m}_b=(0,1,0)$, and $\vec{m}_c=(-\lambda,0,1)/\sqrt{1+\lambda^2}$. The observable associated with $\vec{m}_a$ reveals the exceptional contribution proportional to $n\lambda^{n-1}$, producing a characteristic nonmonotonic transient. In contrast, the observables associated with $\vec{m}_b$ and $\vec{m}_c$ satisfy $m_x+\lambda m_z=0$ and are therefore blind to the exceptional contribution, displaying purely exponential relaxation toward $p_{+}=1/2$. For this simulation we considered $\lambda = 0.8$.}
	\label{fig:visible_blind}
\end{figure}

Figure \ref{fig:visible_blind} shows how the visibility of the exceptional point depends
on the observable, even though the channel and the initial state are kept fixed. For the direction $\vec{m}_a=(1,0,0)$, Eq. \eqref{pplus} reduces to
\begin{equation}
	p_+^{(a)}(n)
	=
	\frac{1}{2}
	\left[
	1+
	\frac{1-\lambda^2}{\sqrt{2}(1+\lambda^2)}
	n\lambda^{n-1}
	\right].
\end{equation}
In this case, the probability contains only the polynomial--exponential
contribution generated by the Jordan block. The term $n\lambda^{n-1}$ is therefore a direct operational signature of the exceptional
structure of the channel.

For the direction $\vec{m}_b=(0,1,0)$, on the other hand,
$m_x+\lambda m_z=0$, and the exceptional contribution vanishes. We obtain
\begin{equation}
	p_+^{(b)}(n)
	=
	\frac{1}{2}
	\left(
	1+\frac{\lambda^{2n}}{\sqrt{2}}
	\right),
\end{equation}
which displays a simple exponential relaxation. This result could be
expected since $\sigma_y$ belongs to a sector decoupled from the Jordan
block. More interestingly, the same absence of an exceptional signature
occurs for
$\vec{m}_c=(-\lambda,0,1)/\sqrt{1+\lambda^2}$. In this case,
\begin{equation}
	p_+^{(c)}(n)
	=
	\frac{1}{2}
	\left[
	1+
	\frac{\lambda^n}{\sqrt{2(1+\lambda^2)}}
	\right].
\end{equation}
Although this observable probes the same $x$-$z$ plane in which the
Jordan block acts, its measurement direction satisfies
$m_x+\lambda m_z=0$. Equivalently, its elementary effects have vanishing
overlap with the eigenoperator
$X_0=\sigma_x+\lambda\sigma_z$ responsible for the polynomial term.
The exceptional contribution is therefore completely suppressed.

These results illustrate the distinction between the algebraic
existence of an exceptional point and its operational visibility.
The same exceptional channel, acting on the same initial state, may
produce either a characteristic polynomial--exponential transient or
an apparently ordinary exponential relaxation, depending only on the
observable chosen at the output. Since $0<\lambda<1$, all traceless
components eventually decay and the three probabilities approach
$p_+=1/2$, corresponding to the maximally mixed stationary state.

\section{Conclusions}
\label{sec:conclusions}

We have formulated exceptional-point dynamics directly in terms of repeated quantum channels and experimentally accessible probabilities. In this framework, no effective non-Hermitian Hamiltonian, Liouvillian generator, or continuous-time description is required. A nontrivial Jordan block of a CPTP map generates polynomial--exponential contributions under repeated applications of the channel, while the appearance of these terms in measurement statistics depends on the overlap between the chosen preparation, the Jordan directions, and the measured effects. Exceptional behavior is therefore an intrinsic property of the channel, whereas its experimental visibility depends on the complete preparation, transformation and measurement scheme.

These ideas were illustrated with an analytically solvable qubit channel possessing a second-order exceptional point. For an arbitrary initial Bloch vector and a general binary projective observable, we obtained the exact probability distribution after $n$ channel applications and identified the contribution proportional to $n\lambda^{n-1}$. The example shows explicitly that, for the same exceptional channel and the same initial state, one effect may reveal the characteristic exceptional transient while another may completely suppress it. This provides a simple operational characterization of exceptional quantum channels directly at the level of measurable probabilities.

\begin{acknowledgments}
The author acknowledges financial support from the Conselho Nacional de Desenvolvimento Científico e Tecnológico (CNPq).
\end{acknowledgments}

\appendix

\bibliography{operational_exceptional_points}

@article{BenderBoettcher1998,
  author  = {Bender, Carl M. and Boettcher, Stefan},
  title   = {Real Spectra in Non-Hermitian Hamiltonians Having {$\mathcal{PT}$} Symmetry},
  journal = {Phys. Rev. Lett.},
  volume  = {80},
  pages   = {5243--5246},
  year    = {1998},
  doi     = {10.1103/PhysRevLett.80.5243}
}

@article{Heiss2012,
  author  = {Heiss, W. D.},
  title   = {The Physics of Exceptional Points},
  journal = {J. Phys. A: Math. Theor.},
  volume  = {45},
  number  = {44},
  pages   = {444016},
  year    = {2012},
  doi     = {10.1088/1751-8113/45/44/444016}
}

@article{ElGanainy2018,
  author  = {El-Ganainy, Ramy and Makris, Konstantinos G. and Khajavikhan, Mercedeh and Musslimani, Ziad H. and Rotter, Stefan and Christodoulides, Demetrios N.},
  title   = {Non-Hermitian Physics and {PT} Symmetry},
  journal = {Nat. Phys.},
  volume  = {14},
  pages   = {11--19},
  year    = {2018},
  doi     = {10.1038/nphys4323}
}

@article{Ashida2020,
  author  = {Ashida, Yuto and Gong, Zongping and Ueda, Masahito},
  title   = {Non-Hermitian Physics},
  journal = {Adv. Phys.},
  volume  = {69},
  pages   = {249--435},
  year    = {2020}
}

@article{Minganti2019,
  author  = {Minganti, Fabrizio and Miranowicz, Adam and Chhajlany, Ravindra W. and Nori, Franco},
  title   = {Quantum Exceptional Points of Non-Hermitian Hamiltonians and Liouvillians: The Effects of Quantum Jumps},
  journal = {Phys. Rev. A},
  volume  = {100},
  pages   = {062131},
  year    = {2019},
  doi     = {10.1103/PhysRevA.100.062131}
}

@article{Minganti2020,
  author  = {Minganti, Fabrizio and Miranowicz, Adam and Chhajlany, Ravindra W. and Arkhipov, Ievgen I. and Nori, Franco},
  title   = {Hybrid-Liouvillian Formalism Connecting Exceptional Points of Non-Hermitian Hamiltonians and Liouvillians via Postselection of Quantum Trajectories},
  journal = {Phys. Rev. A},
  volume  = {101},
  pages   = {062112},
  year    = {2020},
  doi     = {10.1103/PhysRevA.101.062112}
}

@article{Khandelwal2021,
  author  = {Khandelwal, Shishir and Brunner, Nicolas and Haack, G\'eraldine},
  title   = {Signatures of Liouvillian Exceptional Points in a Quantum Thermal Machine},
  journal = {PRX Quantum},
  volume  = {2},
  pages   = {040346},
  year    = {2021},
  doi     = {10.1103/PRXQuantum.2.040346}
}

@article{Chen2022,
  author  = {Chen, Weijian and Abbasi, Maryam and Ha, Byung and Erdamar, Serra and Joglekar, Yogesh N. and Murch, Kater W.},
  title   = {Decoherence-Induced Exceptional Points in a Dissipative Superconducting Qubit},
  journal = {Phys. Rev. Lett.},
  volume  = {128},
  pages   = {110402},
  year    = {2022},
  doi     = {10.1103/PhysRevLett.128.110402}
}

@article{SunYi2024,
  author  = {Sun, Konghao and Yi, Wei},
  title   = {Encircling the Liouvillian Exceptional Points: A Brief Review},
  journal = {AAPPS Bull.},
  volume  = {34},
  pages   = {22},
  year    = {2024},
  doi     = {10.1007/s43673-024-00129-3}
}

@article{WongZengLi2026,
  author  = {Wong, Wai Chun and Zeng, Bei and Li, Jensen},
  title   = {Non-Markovian Exceptional Points by Interpolating Quantum Channels},
  journal = {npj Quantum Inf.},
  volume  = {12},
  pages   = {63},
  year    = {2026},
  doi     = {10.1038/s41534-026-01205-2}
}

@article{PopkovSalerno2026,
  author  = {Popkov, Vladislav and Salerno, Mario},
  title   = {Liouvillian Exceptional Points in Quantum Brickwork Circuits},
  journal = {Phys. Rev. A},
  volume  = {113},
  pages   = {043525},
  year    = {2026},
  doi     = {10.1103/ksj2-w5cj}
}

@book{HeinosaariZiman2012,
  author    = {Heinosaari, Teiko and Ziman, M\'ario},
  title     = {The Mathematical Language of Quantum Theory: From Uncertainty to Entanglement},
  publisher = {Cambridge University Press},
  address   = {Cambridge},
  year      = {2012},
  doi       = {10.1017/CBO9781139031103}
}

@book{Kraus1983,
  author    = {Kraus, Karl},
  title     = {States, Effects, and Operations: Fundamental Notions of Quantum Theory},
  series    = {Lecture Notes in Physics},
  volume    = {190},
  publisher = {Springer},
  address   = {Berlin},
  year      = {1983},
  doi       = {10.1007/3-540-12732-1}
}

@book{NielsenChuang2010,
  author    = {Nielsen, Michael A. and Chuang, Isaac L.},
  title     = {Quantum Computation and Quantum Information},
  edition   = {10th Anniversary Edition},
  publisher = {Cambridge University Press},
  address   = {Cambridge},
  year      = {2010}
}

@article{Choi1975,
  author  = {Choi, Man-Duen},
  title   = {Completely Positive Linear Maps on Complex Matrices},
  journal = {Linear Algebra Appl.},
  volume  = {10},
  number  = {3},
  pages   = {285--290},
  year    = {1975},
  doi     = {10.1016/0024-3795(75)90075-0}
}

@article{EvansHoeghKrohn1978,
  author  = {Evans, David E. and H\o egh-Krohn, Raphael},
  title   = {Spectral Properties of Positive Maps on {$C^*$}-Algebras},
  journal = {J. Lond. Math. Soc.},
  volume  = {s2-17},
  number  = {2},
  pages   = {345--355},
  year    = {1978},
  doi     = {10.1112/jlms/s2-17.2.345}
}

@article{HansonRouzeFranca2020,
  author  = {Hanson, Eric P. and Rouz\'e, Cambyse and Fran\c{c}a, Daniel Stilck},
  title   = {Eventually Entanglement Breaking Markovian Dynamics: Structure and Characteristic Times},
  journal = {Ann. Henri Poincar\'e},
  volume  = {21},
  pages   = {1517--1571},
  year    = {2020},
  doi     = {10.1007/s00023-020-00906-4}
}

@article{MiriAlu2019,
	author  = {Miri, Mohammad-Ali and Al{\`u}, Andrea},
	title   = {Exceptional points in optics and photonics},
	journal = {Science},
	volume  = {363},
	number  = {6422},
	pages   = {eaar7709},
	year    = {2019},
	doi     = {10.1126/science.aar7709}
}

@article{Ozdemir2019,
	author  = {{\"O}zdemir, {\c{S}}ahin K. and Rotter, Stefan and Nori, Franco and Yang, Lan},
	title   = {Parity--time symmetry and exceptional points in photonics},
	journal = {Nature Materials},
	volume  = {18},
	pages   = {783--798},
	year    = {2019},
	doi     = {10.1038/s41563-019-0304-9}
}

@article{Bergholtz2021,
	author  = {Bergholtz, Emil J. and Budich, Jan Carl and Kunst, Flore K.},
	title   = {Exceptional topology of non-Hermitian systems},
	journal = {Reviews of Modern Physics},
	volume  = {93},
	pages   = {015005},
	year    = {2021},
	doi     = {10.1103/RevModPhys.93.015005}
}

@article{LiEtAl2023,
	author  = {Li, Aodong and Wei, Heng and Cotrufo, Michele and Chen, Weijin and Mann, Sander and Ni, Xiang and Xu, Bingcong and Chen, Jianfeng and Wang, Jian and Fan, Shanhui and Qiu, Cheng-Wei and Al{\`u}, Andrea and Chen, Lin},
	title   = {Exceptional points and non-Hermitian photonics at the nanoscale},
	journal = {Nature Nanotechnology},
	volume  = {18},
	pages   = {706--720},
	year    = {2023},
	doi     = {10.1038/s41565-023-01408-0}
}

@article{Naghiloo2019,
	author  = {Naghiloo, M. and Abbasi, M. and Joglekar, Yogesh N. and Murch, K. W.},
	title   = {Quantum state tomography across the exceptional point in a single dissipative qubit},
	journal = {Nature Physics},
	volume  = {15},
	pages   = {1232--1236},
	year    = {2019},
	doi     = {10.1038/s41567-019-0652-z}
}

@article{Erdamar2024,
	author  = {Erdamar, Serra and Abbasi, Maryam and Ha, Byung and Chen, Weijian and Muldoon, Jacob and Joglekar, Yogesh and Murch, Kater W.},
	title   = {Constraining work fluctuations of non-Hermitian dynamics across the exceptional point of a superconducting qubit},
	journal = {Physical Review Research},
	volume  = {6},
	pages   = {L022013},
	year    = {2024},
	doi     = {10.1103/PhysRevResearch.6.L022013}
}

@article{Abo2024,
	author  = {Abo, Shilan and Tulewicz, Patrycja and Bartkiewicz, Karol and {\"O}zdemir, {\c{S}}ahin K. and Miranowicz, Adam},
	title   = {Experimental Liouvillian exceptional points in a quantum system without Hamiltonian singularities},
	journal = {New Journal of Physics},
	volume  = {26},
	number  = {12},
	pages   = {123032},
	year    = {2024},
	doi     = {10.1088/1367-2630/ad98b6}
}

@article{Zhang2025,
	author  = {Zhang, Hao-Long and Han, Pei-Rong and Wu, Fan and Ning, Wen and Yang, Zhen-Biao and Zheng, Shi-Biao},
	title   = {Experimental Observation of Non-Markovian Quantum Exceptional Points},
	journal = {Physical Review Letters},
	volume  = {135},
	pages   = {230203},
	year    = {2025},
	doi     = {10.1103/jk6y-55xp}
}

@article{Kopciuch2025,
	author  = {Kopciuch, Marek and Miranowicz, Adam},
	title   = {Liouvillian and Hamiltonian exceptional points of atomic vapors: The spectral signatures of quantum jumps},
	journal = {Physical Review Research},
	volume  = {7},
	pages   = {033187},
	year    = {2025},
	doi     = {10.1103/zxw5-nlsn}
}

@article{Ciccarello2022,
	author  = {Ciccarello, Francesco and Lorenzo, Salvatore and Giovannetti, Vittorio and Palma, G. Massimo},
	title   = {Quantum collision models: Open system dynamics from repeated interactions},
	journal = {Physics Reports},
	volume  = {954},
	pages   = {1--70},
	year    = {2022},
	doi     = {10.1016/j.physrep.2022.01.001}
}

@article{AttalPautrat2006,
	author  = {Attal, St{\'e}phane and Pautrat, Yan},
	title   = {From repeated to continuous quantum interactions},
	journal = {Annales Henri Poincar{\'e}},
	volume  = {7},
	number  = {1},
	pages   = {59--104},
	year    = {2006},
	doi     = {10.1007/s00023-005-0242-8}
}

\end{document}